\documentclass[5p,times]{elsarticle}
\usepackage{multicol}
\usepackage{amsmath}
\usepackage{graphicx}
\usepackage{subcaption}
\usepackage[colorlinks=true, allcolors=blue]{hyperref}
\usepackage{xcolor}
\usepackage{amsmath,amssymb,amsfonts}
\usepackage{import}
\usepackage{qcircuit}
\usepackage{braket}
\usepackage{adjustbox}
\usepackage{graphicx}
\usepackage{adjustbox}
\usepackage{algorithm}
\usepackage{multirow}
\usepackage{algpseudocode} 
\usepackage{balance}
\newcommand\scalemath[2]{\scalebox{#1}{\mbox{\ensuremath{\displaystyle #2}}}}

\journal{Future Generation Computer Systems}
\begin{document}
\begin{frontmatter}

\title{LuGo: an Enhanced Quantum Phase Estimation Implementation}

\author[nccs]{Chao Lu\corref{cor1}}
\author[nccs]{Muralikrishnan Gopalakrishnan Meena}
\author[nccs]{Kalyana C. Gottiparthi}

\address[nccs]{
    National Center for Computational Sciences, 
    Oak Ridge National Laboratory,
    Oak Ridge, TN, USA
}

\cortext[cor1]{To whom correspondence should be addressed: \href{luc1@ornl.gov}{luc1@ornl.gov}.}

\cortext[cor2]{\scriptsize This manuscript has been authored by UT-Battelle, LLC, under contract DE-AC05-00OR22725 with the US Department of Energy (DOE). The publisher acknowledges the US government license to provide public access under the DOE Public Access Plan (\url{http://energy.gov/downloads/doe-public-access-plan}).}

\begin{abstract}
Quantum Phase Estimation (QPE) is a cardinal algorithm in quantum computing that plays a crucial role in various applications, including cryptography, molecular simulation, and solving systems of linear equations. However, the standard implementation of QPE faces challenges related to time complexity and circuit depth, which limit its practicality for large-scale computations. We introduce LuGo, a novel framework designed to enhance the performance of QPE by reducing circuit duplication, as well as using parallelization techniques to achieve faster generation of the QPE circuit and gate reduction. We validate the effectiveness of our framework by generating quantum linear solver circuits, which require both QPE and inverse QPE, to solve linear systems of equations. LuGo achieves significant improvements in both computational efficiency and hardware requirements without compromising on accuracy. Compared to a standard QPE implementation, LuGo reduces time consumption to generate a circuit that solves a $2^6\times 2^6$ system matrix by a factor of $50.68$ and over $31\times$ reduction of quantum gates and circuit depth, with no fidelity loss on an ideal quantum simulator. We demonstrated the versatility and scalability of LuGo enabled HHL algorithm by simulating a canonical Hele-Shaw fluid problem using a quantum simulator. With these advantages, LuGo paves the way for more efficient implementations of QPE, enabling broader applications across several quantum computing domains.
\end{abstract}
\end{frontmatter}

\section{Introduction} \label{sec:intro}
Quantum computing has demonstrated significant promise through advancements in algorithms and hardware. Leading technology companies are exploring diverse approaches to develop robust, large-scale quantum computers capable of implementing these algorithms. Notably, IBM now provides public access to superconducting quantum computers with over 150 qubits via its cloud platform~\cite{qiskit}. Google's Sycamore quantum chip has successfully implemented an efficient quantum error-correction code, achieving an error rate below $10^{-6}$ \cite{googleecc}. Quantum algorithms are proving to offer exponential speedups over their classical counterparts in areas such as machine learning, cryptography, quantum physics, and fluid dynamics \cite{machinelearning,li2023international,di2022dawn, shor,physicsim,gopalakrishnan2024solving,meena2024towards,lapworthHHL,ye2024hybridcfd,dodin2021applications}. 

However, considerable efforts are still needed to enable quantum computers to effectively address real-world problems. In parallel, the development of quantum algorithms remains critical for efficient design and implementation. Several groundbreaking algorithms, including Shor's algorithm, Grover's search algorithm, the Quantum Approximate Optimization Algorithm (QAOA), the Quantum Linear Systems  Algorithm (QLSA), and the Quantum Phase Estimation (QPE) algorithm have been proposed to tackle computational challenges that are infeasible for classical systems~\cite{quantumbook,montanaro2016quantum}. Despite their theoretical potential for exponential acceleration, the practical implementation and scalability of many of these algorithms face significant hurdles.

Quantum Phase Estimation (QPE), as a fundamental algorithm in quantum computing, serves as a cornerstone for various applications. It was first introduced in 1995 to estimate the phase of a unitary matrix \cite{kitaev1995quantum}.
Over the years, numerous studies have focused on enhancing the implementation and optimization of QPE to improve its efficiency and scalability. 
Among these, QPE stands out for its versatility in solving problems with exponential speedups. As one of the cornerstone algorithms in quantum computation, QPE has already been integrated in several quantum software development kits~\cite{qiskit,cirq,bergholm2018pennylane,cudaq,braket}. However, QPE faces several challenges, including generating controlled-unitary circuits, which are critical to its functionality. Optimizing QPE design is crucial to improve the performance of various quantum algorithms that rely on it \cite{quantumbook}. Moreover, maintaining high fidelity while executing QPE on real quantum hardware remains difficult. 

\begin{figure*}[bt]
    \centering
    \includegraphics[width=1.1\linewidth]{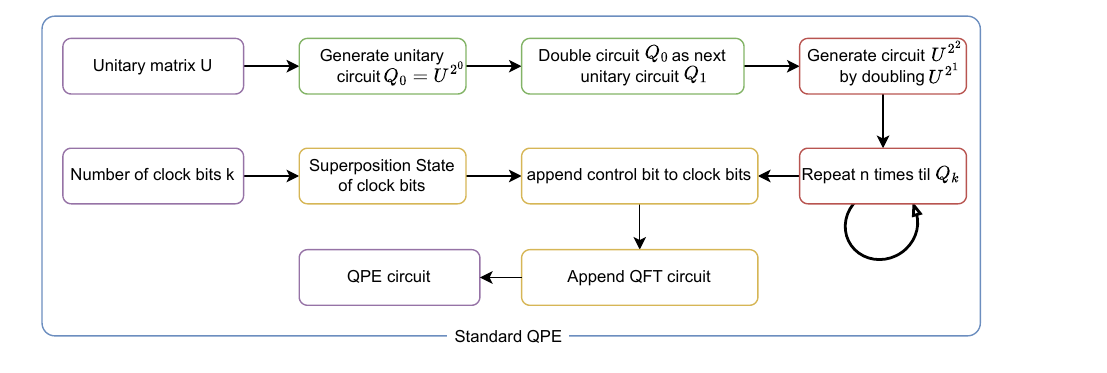}
    \caption{Standard QPE circuit generation process.}
    \label{fig:standardQPE}
\end{figure*}

Several directions for improving the fidelity of QPE have been attempted for more robust and efficient implementation. An evolutionary algorithm was applied to reduce the fidelity loss due to Trotter Approximation for QPE algorithm \cite{daspal2024minimizing}. Kang et al. optimized QPE algorithm specifically for simulating electronic states, introducing `QPESIM' for the simulation \cite{kang2022optimized}. Iterative QPE implementations have also been explored for better precision \cite{ni2023low,smith2022iterative,van2020iterative}. Li proposed an iterative QPE implementation for precise readout on noisy intermediate-scale quantum (NISQ) devices with only a few ancilla qubits~\cite{li2024iterative}. Van et al. improved iterative QPE algorithm for optimized sample complexity, enabling minimal sampling size required for acceptable accuracy \cite{van2020iterative}. Smith et al. \cite{smith2022iterative} proposed an iterative QPE for reduced circuit implementation. Mohammadbagherpoor et al. improved QPE circuits by removing unnecessary controlled unitary gates and replaced with unitary operators \cite{mohammadbagherpoor2019improved}. However, such implementations are applicable only in limited scenarios where the controlled unitary operators can be replaced. Others, use alternative approaches to replace part of the QPE algorithm, where it either does not generalize well to most quantum hardware \cite{johnstun2021optimizing}, reduce accuracy \cite{rall2021faster}, or require additional circuit complexity \cite{dong2024optimal}.

To extend the capability of quantum algorithms that depends on QPE, improving QPE is an opportunity to pave the way for practical quantum computing solutions that could redefine our understanding of the world and solve problems once thought insurmountable. We introduce LuGo, a QPE circuit design aiming for a scalable and efficient generation of the quantum circuit. We demonstrate the capability of LuGo on a quantum linear solver algorithm (QLSA), the Harrow--Hassidim--Lloyd (HHL) algorithm~\cite{HHL}. The major contributions of this work are:
\begin{itemize}
    \item We introduce LuGo, a new Quantum Phase Estimation (QPE) circuits generation approach for circuit generation time and quantum gate reduction.
    \item We applied LuGo on the HHL algorithm that utilize QPE as a sub-algorithm to illustrate the efficiency of LuGo.
    \item We validate and benchmark the new HHL circuit generation algorithm using ideal simulator run on the Frontier supercomputer to analyze the scalability and fidelity of the proposed LuGo framework.
    \item LuGo achieves circuit generation time consumption reduction by $50.68\times$ and over $31\times$ on circuit depth and gate reduction for a $2^6\times 2^6$ matrix input. When transpiled to basis gates `u3' and `cx', LuGo obtained $37.07\times$ and $40.97\times$ reduction of `u3' and `cx' gates, respectively.
    \item LuGo does not suffer from fidelity loss on an ideal quantum simulator, achieving identical fidelity performance with the standard QPE approach.
    \item We performed a canonical Hele-Shaw fluid flow problem using LuGo-enabled HHL algorithm, demonstrating versatility and better scalability over the standard approach.
\end{itemize}

The rest of this paper is organized as follows.
Section~\ref{sec:background} provides background and methodology, including the background information of QPE algorithm, the proposed LuGo framework, and the HHL algorithm which we use to demonstrate the capability of LuGo. Then, Section~\ref{sec:Results} describes performance comparison of LuGo and standard QPE with respect to time consumption, circuit statistics, memory footprint, and fidelity as integrated in HHL circuits and run on ideal quantum simulator. Lastly, we provide some discussion and concluding remarks in Section~\ref{sec:Conclusion}.

 \section{Methods}\label{sec:background}
\subsection{Quantum Phase Estimation (QPE)}\label{sec:QPE}

\begin{figure}[htb!]

\[
\scalemath{0.86}{
\Qcircuit @C=1em @R=1em {
    \lstick{\ket{0}} & \gate{H} & \ctrl{4} & \qw &\qw    & \qw    & \push{\cdots} \qw & \multigate{3}{\mathrm{QFT}^\dagger} & \meter & \cw & \rstick{\ket{\lambda_1}}\\
    \lstick{\ket{0}} & \gate{H} & \qw      & \ctrl{3} & \qw      & \qw  &\push{\cdots} \qw  & \ghost{\mathrm{QFT}^\dagger}& \meter & \cw & \rstick{\ket{\lambda_2}}\\
    \lstick{\vdots}  &          &          & &&   & \vdots \\
    \lstick{\ket{0}} & \gate{H} & \qw      & \qw      & \qw   & \ctrl{1} &\push{ \cdots} \qw & \ghost{\mathrm{QFT}^\dagger} & \meter &\cw & \rstick{\ket{\lambda_k}}\\
    \lstick{\ket{\phi}} & \qw & \multigate{3}{U^{2^0}} & \multigate{3}{U^{2^1}} & \qw & \multigate{3}{U^{2^k}} & \push{\cdots} \qw & \qw & \meter &\cw & \rstick{\ket{\phi}}\\
    \lstick{\ket{\phi}} & \qw      & \ghost{U^{2^0}} & \ghost{U^{2^0}} & \qw    & \ghost{U^{2^0}} & \push{\cdots}  \qw     & \qw &\meter& \cw&\rstick{\ket{\phi}}\\
    \lstick{\vdots}  &          &          &                                    &&   & \vdots \\
    \lstick{\ket{\phi}} & \qw      & \ghost{U^{2^0}} & \ghost{U^{2^0}} & \qw    & \ghost{U^{2^0}} & \push{\cdots} \qw &\qw   & \meter & \cw &\rstick{\ket{\phi}}\\ }}
    \]

    \caption{QPE circuit to estimate phase of input $u$. The QPE algorithm will estimate the state of $\ket{u}$ and generate eigenvalues $\lambda$ of matrix $u$}
    \label{fig:qpe_circuit}
\end{figure}

\begin{figure*}
\centering
\includegraphics[width=\linewidth]{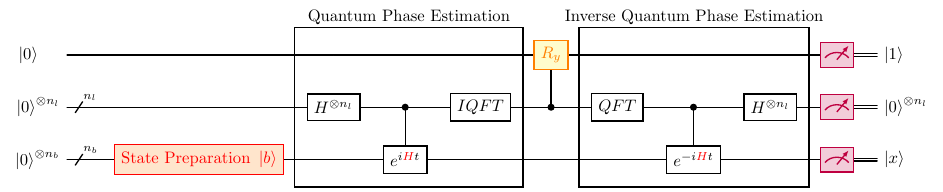}
\caption{Quantum circuit for the HHL algorithm. The circuit includes the QPE part, eigenvalue reciprocal part, the conditional rotation part, and the inverse QPE part. Measurement of the ancilla qubit determines the success of the algorithm.}
    \label{fig:hhl_circuit}
\end{figure*}

QPE is designed to estimate the phase $\ket{\lambda}$ associated with an eigenvector $\ket{\phi}$ of a unitary operator $U$. Fig.~\ref{fig:qpe_circuit} and \ref{fig:standardQPE} depict the overall QPE circuit and circuit generation process, respectively. The circuit consists of two main registers: the clock register and the input qubits. Initially, the clock register, composed of $k$ qubits, is set to the state $\ket{0}$. A Hadamard gate is applied to each qubit in this register, transforming it into a superposition state $\frac{1}{\sqrt{2^k}} \sum_{j=0}^{2^k-1} \ket{j}$, which prepares the clock qubits for phase estimation.

Following the Hadamard operations, controlled-unitary ($\text{c-}U$) gates are applied. Each clock qubit controls a $U$ raised to an exponentially increasing power—specifically, the $j$-th qubit controls $U^{2^{j-1}}$. The novelty of our LuGo implementation is to address this exponentially increasing computation of the controlled-unitary operations, as elaborate in Section~\ref{sec:LuGo}. This sequence of operations entangles the clock register with the input state $\ket{\phi}$, encoding the phase $\phi$ into the clock qubits through the mechanism known as ``phase kickback''. As a result, the combined state becomes $\frac{1}{\sqrt{2^k}} \sum_{j=0}^{2^k-1} e^{2\pi i j \theta} \ket{j} \ket{\phi}$, where $j$ is the dummy index and $k$ is the number of clock qubits.

To decode this encoded phase information, an inverse Quantum Fourier Transform (QFT) is applied to the clock register. This transformation converts the superposition state into a binary representation of $\ket{\lambda}$, enabling us to measure the clock qubits and obtain an estimate of $\ket{\lambda}$ with high precision. Equation~\ref{equ:qpe} depicts the overall procedure:
\begin{equation} \label{equ:qpe}
\left( \frac{1}{\sqrt{2^k}} \sum_{j=0}^{2^k-1} \ket{j} \otimes \ket{\phi} \right)
\xrightarrow{\text{c-}U}
\frac{1}{\sqrt{2^k}} \sum_{j=0}^{2^k-1} e^{2\pi i j \theta} \ket{j} \otimes \ket{\phi}
\xrightarrow{\text{QFT}^\dagger}
\ket{\tilde{\lambda}} \otimes \ket{\phi}.
\end{equation}
Throughout these operations, the input state $\ket{\phi}$ remains unchanged except for acquiring the phase factor $e^{2\pi i j \lambda}$, ensuring that it is preserved for potential further use in subsequent quantum computations.

To design the controlled unitary operations for QPE circuits, an algorithm that encode the unitary matrix to the corresponding quantum circuits is required. The default circuit generation approach in Qiskit (version 1.3) utilizes the Peter--Weyl decomposition algorithm \cite{weyl1, weyl2, weyl3}. Specifically, we use the Qiskit function {\tt control} which automatically triggers the decomposition operation when appending a control point to the customized unitary gate. The Peter--Weyl algorithm decomposes a unitary matrix into a series of controlled-unitary gates, which can be represented as quantum circuits. This approach allows for efficient circuit generation, enabling the implementation of QPE with arbitrary unitary matrices on quantum hardware.

\subsection{Harrow--Hassidim--Lloyd Algorithm} \label{sec:HHL}
The Harrow--Hassidim--Lloyd (HHL) algorithm is a canonical quantum algorithm to solve linear systems of equations~\cite{HHL}. It is one of the quantum algorithms that shows exponential acceleration over classical computation that utilizes QPE as a sub-algorithm for the computation. For the overall computational scheme for the HHL algorithm to solve a linear system of equations $A\vec{x}=\vec{b}$, three parts are required: QPE, reciprocal of eigenvalues, and inverse QPE, as shown in Fig.~\ref{fig:hhl_circuit}. In order to convert a matrix to a unitary matrix for QPE implementation, the input matrix $A$ is initialized as a series of controlled-unitary gates, $e^{iH2^t}$, where $H$ is the input matrix $A$ if it is Hermitian. If the matrix $A$ is not Hermitian, it can be padded and converted to a Hermitian $H = \begin{bmatrix}
A & 0\\
0 & A^\dagger
\end{bmatrix}$, and the vector $b$ can be initialized as $c = \begin{bmatrix}
0\\
b
\end{bmatrix}$ to match matrix $H$. Then, the unitary matrix $e^{iH2^t}$ is computed by a unitary circuits generation algorithm. As discussed in Section~\ref{sec:QPE}, by default in Qiskit, the unitary circuits generation algorithm is performed using the Peter--Weyl decomposition to generate each controlled-unitary circuit. After these circuits are generated, the algorithm constructs reciprocal eigenvalue circuits. Finally, it integrates the controlled-unitary gates with other components of the HHL circuit to complete the circuit generation process.

\subsection{LuGo: a scalable and efficient QPE implementation}\label{sec:LuGo}
Generation of the QPE circuit predominantly consumes time during the creation of the controlled-unitary circuit, particularly in the repetition of controlled-unitary operations for $U^{2^k}$. Therefore, reducing the time required for these controlled-unitary operations is crucial to minimizing both circuit depth and overall computation time. In standard QPE implementations, quantum circuits for the controlled-unitary operation $U = e^{iH2^j}$ (where $j = 1, 2, 3, \ldots, k$) are designed by first constructing the initial controlled-unitary circuit for $e^{iH}$ using the Peter--Weyl decomposition method. Subsequent circuits for $e^{iH2^j}$ are created by replicating this initial circuit. This process results in the generated first unitary circuit being repeated $2^k - 1$ times, where $k$ denotes the number of clock qubits. However, as the number of clock qubits increases linearly with the qubit count, the repetition of controlled-unitary gates grows exponentially. Consequently, this exponential increase significantly affects both the circuit generation time and the total gate count required, particularly as the system scales to larger qubit sizes. 

\begin{figure*}[bt!]
    \centering
    \includegraphics[width=1\linewidth]{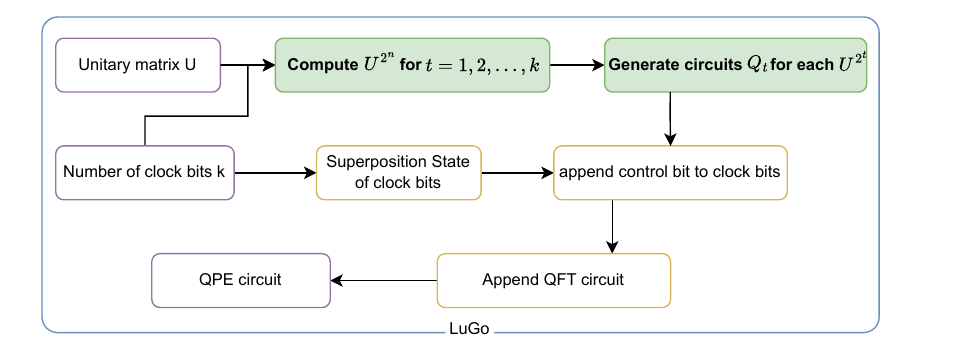}
    \caption{Proposed LuGo designing process. We highlight the operation that obtained acceleration over standard QPE.321}
    \label{fig:LuGo}
\end{figure*}

To address these challenges, we introduce a new approach, namely LuGo, to enhance QPE circuit design and reduce circuit depth without sacrificing fidelity. Our methodology focuses on optimizing the design and execution of QPE by improving the controlled-unitary matrix design efficiency. Fig.~\ref{fig:LuGo} depicts our LuGo framework. Within this framework, each unitary circuit for controlled unitary operations $e^{iH2^j}$ are independently generated using the Peter--Weyl decomposition instead of repeating the initial controlled-unitary circuit for $e^{iH}$. LuGo achieved embarrassingly parallelized computation to eliminate such overhead by generating each controlled unitary circuit independently, leading to a more efficient and scalable QPE circuit generation routine.

\begin{figure}[!bt]
    \begin{algorithm}[H]
    \raggedright\textbf{Input:} Hermitian matrix input $H$, vector input $\ket{\phi}$, and $k$ clock qubits.\\
    \raggedright\textbf{Output:} QPE circuit available to execute on quantum hardware.\\
    \caption{LuGo framework for QPE}
    \label{alg:QPE}
    \begin{algorithmic}[1]
    \For {$t \gets (0$ to $k$)} \textbf{in parallel}
    \State Compute and store unitary matrix $e^{iH2^t}$ as $U_t$.
    \State Generate controlled unitary circuit $Q_t$ based on $U_t$.
\EndFor
    \State Initialize the clock qubits $\ket{k}$ using Hadamard gates.
    \State Prepare the quantum State $\ket {\phi}$ at input qubits.
    \State Append each $Q_t$ to input qubits and clock qubits.
    \State Append $iQFT$ circuit to clock qubits.
    \State \Return Generated QPE circuit.
    \end{algorithmic}
    \end{algorithm}
    \vspace{-5mm}
    \end{figure}

To generate the QPE circuits, LuGo enables the parallel computation of controlled-unitary circuits generation shown in Alg.~\ref{alg:QPE}. Firstly, the QPE circuit takes the Hermitian matrix $H$, input vector $\ket{\phi}$, and number of clock qubits $k$. Once all inputs are fed into the algorithm, the algorithm will first perform the parallel computation for controlled-unitary circuits generation by directly computing the matrix exponent (Lines 1--4). Once the generation of each controlled-unitary circuit finishes, the rest of the procedure is to assemble the components of QPE circuits together by first appending Hadamard gates to clock qubits to initialize clock qubits (Line 5). Next, a quantum state preparation algorithm is performed to initialize the quantum state $\ket{\phi}$ (Line 6). Then, the algorithm will assign each controlled-unitary circuit to clock qubits and input qubits (Line 7). After the inverse QFT circuits are appended to the circuits (Line 8), the algorithm returns the generated QPE circuit as the output (Line 9).

The core advantage introduced by LuGo is to reduce the complexity of generating controlled unitary circuits of $e^{iHj}$ when $j$ is large. For the standard QPE generation algorithm, the complexity to generate the controlled unitary circuits is linear concerning $j$. However, classical algorithms for the matrix exponential can accelerate the computation with \textbf{logarithmic complexity} with respect to $j$ scales~\cite{expm}. 
By employing the classical matrix exponential algorithm, each controlled-unitary circuit can be generated independently. This independence allows for the distribution of computational tasks across multiple processing units, enabling efficient parallel computation. The number of clock qubits determines the maximum number of CPU cores required by LuGo. Consequently, the time required to generate the QPE circuit is significantly reduced, as different segments of the circuit can be generated simultaneously without sequential dependencies.

For scenarios where designing controlled-unitary circuits for specific unitary matrices is challenging for certain unitary circuit generation algorithms, the standard approach can override the LuGo approach for the unitary matrix $U^{2^j}$ by duplicating the unitary matrix $U^{2^{j-1}}$ to perform the computation.

\subsection{Complexity analysis} \label{sec:compleixty}
The number of qubits used for the QPE circuit generation will remain unchanged between the standard QPE and LuGo implementations. The total number of qubits required for a QPE circuit is $n+k$, where $k$ is the clock qubits and $n$ input qubits. The number of clock qubits can be written as $k = \mathcal{O}(\log(1/\epsilon))$, where $\epsilon$ is the precision of the QPE algorithm. For quantum gate complexity, the standard QPE requires the design of the first controlled-unitary circuit and repeats $2^k$ times for the computation of  $U^{2^k}$. Assume the design of the first controlled unitary circuit with complexity of $\mathcal{C}(U)$, and the gates complexity of inserting an inverse QFT circuit is $\mathcal{O}(k\cdot \log k)$ for the optimized approach. Note that the complexity of QFT is much less than the controlled unitary gates. These lead to the gate complexity of a standard QPE to be $\mathcal{O}(2^k\mathcal{C}(U))$. 

The depth of a quantum circuit is defined as the minimum number of time steps required to execute all of its gates. A single time step corresponds to the simultaneous execution of a set of gates that act on disjoint qubits. Thus, if $N$ gates can be applied in parallel, they collectively contribute only one time step. Conversely, if the $N$ gates must be applied sequentially, they require $N$ time steps, yielding a circuit depth of $N$.
The gate count is the total number of quantum gates in the quantum circuit, including single and double qubits gates. 
Since all the components mentioned above are sequential for the QPE circuit, the circuit depth complexity is similar to the gate complexity, which is $\mathcal{O}(2^k\mathcal{D}(U))$, where $\mathcal{D}(U)$ denotes the depth complexity of the generated controlled-unitary circuit. For LuGo, the controlled-unitary circuit is generated independently, which means it is not formed by the repetition of the first controlled-unitary gate. Therefore, the QPE circuit generation process does not require exponential stacking of controlled-unitary gates. The gate complexity of LuGo, therefore, is $\mathcal{O}(k\mathcal{C}(U))$. Thus, the generation of quantum circuits is throttled by memory consumption, which is related to the number of qubits and the number of quantum gates. 

The LuGo algorithm introduced here offers significant advantages by reducing overall computation time. The computation of $e^{iH2^j}$ is optimized using an efficient matrix exponential algorithm on a classical computer, which reduces computational overhead on the quantum side~\cite{expm}. Classical computation for matrix exponential by squaring, a well-studied approach, has a time complexity of $\mathcal{O}(\log n)$ concerning the exponent indexes, making it far more efficient than the linear stack of controlled-unitary circuits on the standard QPE side. This optimization results in substantial reductions in both time and gate count for QPE circuit generation.
 \section{Results} \label{sec:Results}
In this section, we evaluate the performance of our LuGo framework compared with the standard QPE circuit generation approach. The QPE implementations are demonstrated on the HHL algorithm \cite{HHL}, a canonical QLSA algorithm. All developments are performed using Qiskit (version 1.3) \cite{qiskit}. We utilize tridiagonal Toeplitz matrices as the linear systems of equations to benchmark the two QPE circuits generation approaches. However, LuGo and standard QPE can take any arbitrary unitary matrix as the input to solve linear systems of equations, and any matrix can be transformed to a unitary matrix as discussed in Section \ref{sec:HHL}.

\begin{figure*}[bt!]
 \begin{subfigure}{.49\textwidth}
   \centering
   \includegraphics[width=0.8\linewidth]{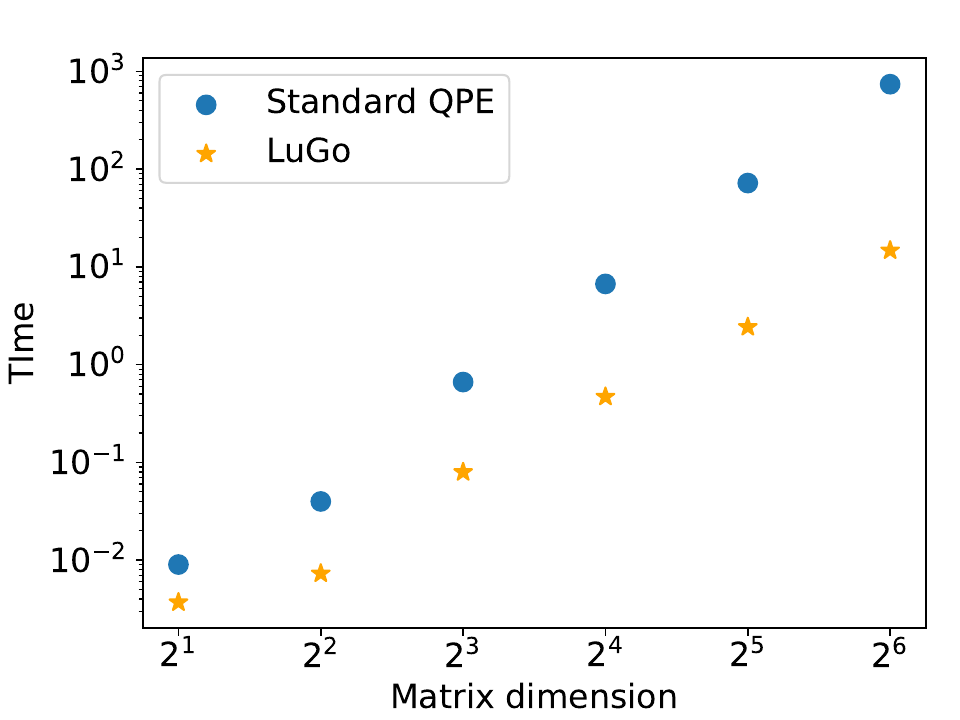}
   \caption{QPE}
   \label{fig:time_qpe}
 \end{subfigure}
 \begin{subfigure}{.49\textwidth}
   \centering
   \includegraphics[width=0.8\linewidth]{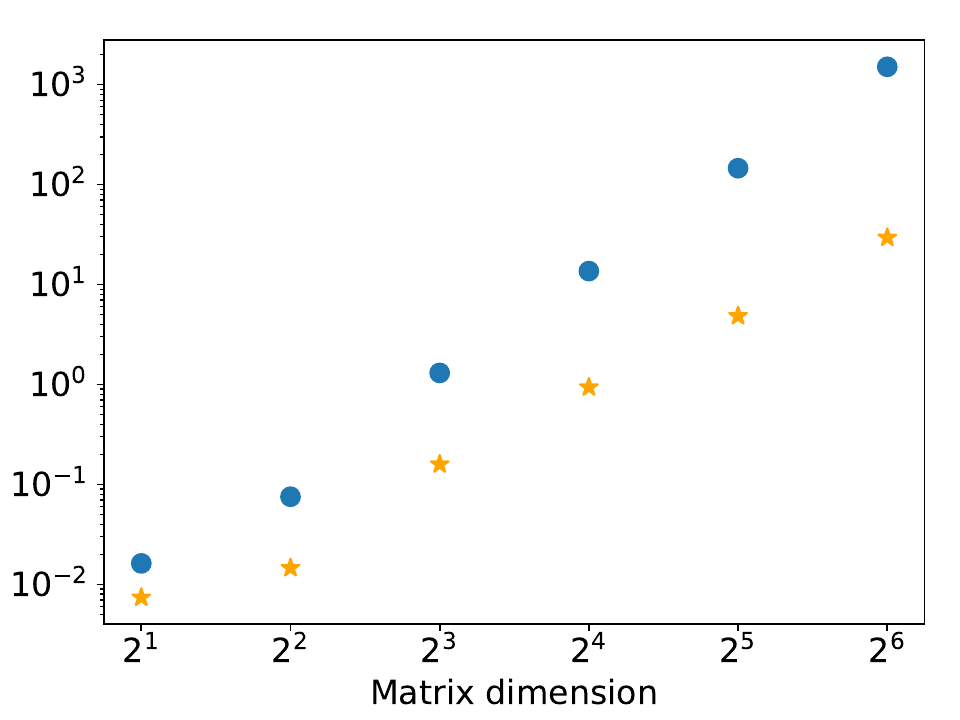}
   \caption{QPE+iQPE}
   \label{fig:time_qpe+iqpe}
 \end{subfigure}
 \begin{subfigure}{.49\textwidth}
  \centering
  \includegraphics[width=0.8\linewidth]{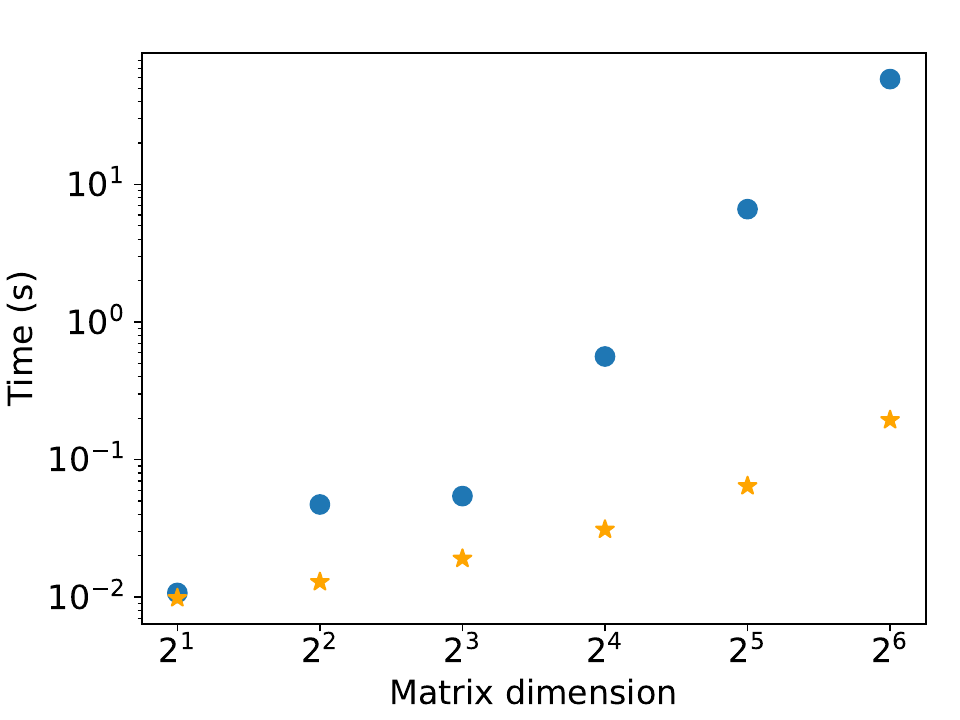}
  \caption{Other Components}
  \label{fig:time_other_components}
   \end{subfigure}
   \begin{subfigure}{.49\textwidth}
  \centering
  \includegraphics[width=0.8\linewidth]{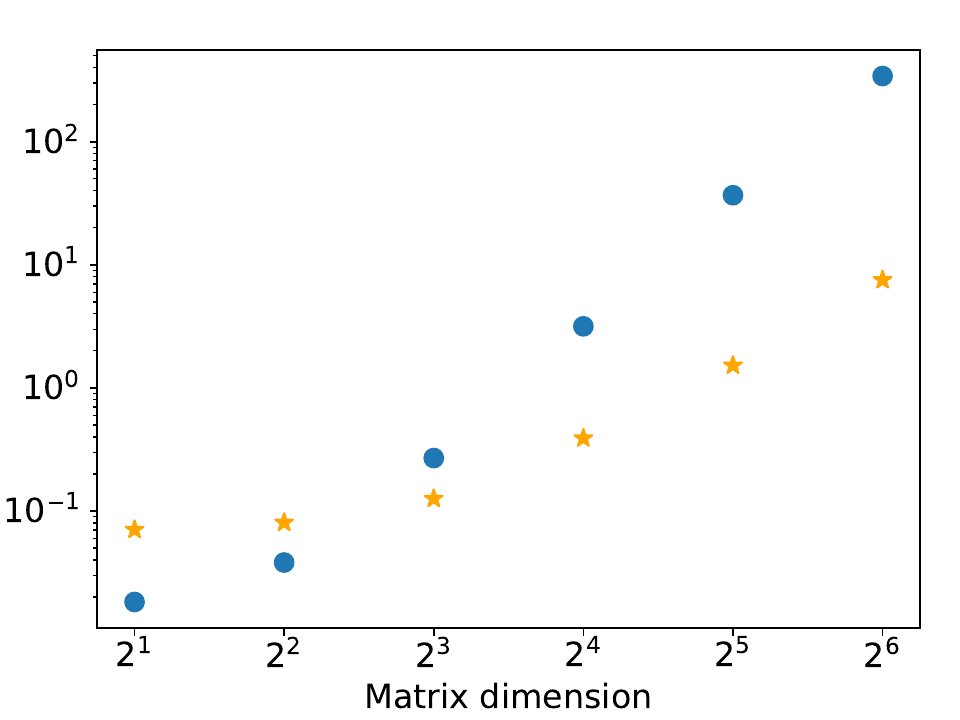}
  \caption{Circuit storage}
  \label{fig:time_save}
   \end{subfigure}  
 \caption{Time to generate HHL circuits using standard QPE and LuGo with single thread version. LuGo demonstrates advantages on circuit generation of QPE and inverse QPE, Other components of QPE circuits, and circuit storage in most of the scenarios.}
 \label{fig:time_circ_gen}
\end{figure*}

\begin{figure}[bt!]
\centering
  \includegraphics[width=0.98\linewidth]{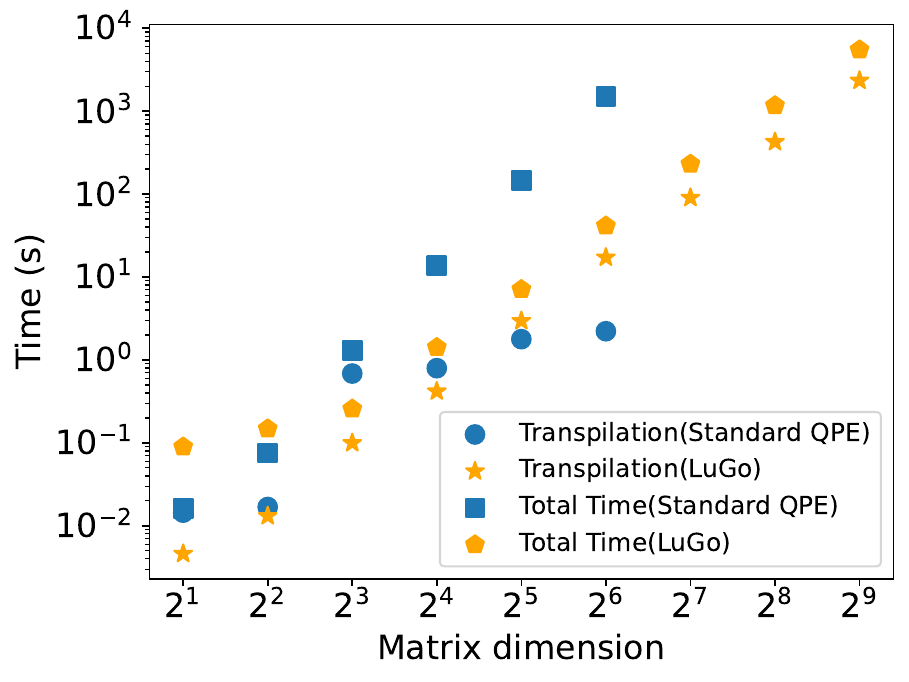}
  \caption{Time consumption for unitary matrix decomposition and total time consumption using standard QPE and LuGo approach.}
  \label{fig:transpilation}
\end{figure}

To measure execution time, we used the Oak Ridge Leadership Computing Facility's Frontier supercomputer \cite{atchley2023frontier} and set a time limit of two hours (Max Walltime for single node reservation on Frontier) to determine the maximum size of the HHL circuit that can be generated within this constraint. For each node in the Frontier supercomputer consists of one 64-core AMD “Optimized 3rd Gen EPYC” CPU (with 2 hardware threads per physical core) with access to 512 GB of DDR4 memory. For each execution, we set timers for each component of the HHL circuit generation procedure, including time consumption of QPE and inverse QPE circuits, summation of various other components of the HHL circuit since they take less than $1\%$ of the entire computation, and to save the circuit in a {\tt qpy} format file (a binary serialization format integrated in Qiskit for {\tt QuantumCircuit} and {\tt ScheduleBlock} objects). These measurements provide detailed insights into the time consumption of each process. Finally, we use Qiskit's AerSimulator (version 0.16.0) to simulate the generated HHL circuits for validating the accuracy of LuGo by comparing the fidelity of the results obtained using LuGo with that from the standard QPE implementation.

\begin{figure*}[bt]
 \begin{subfigure}{.495\textwidth}
   \centering
   \includegraphics[width=0.9\linewidth]{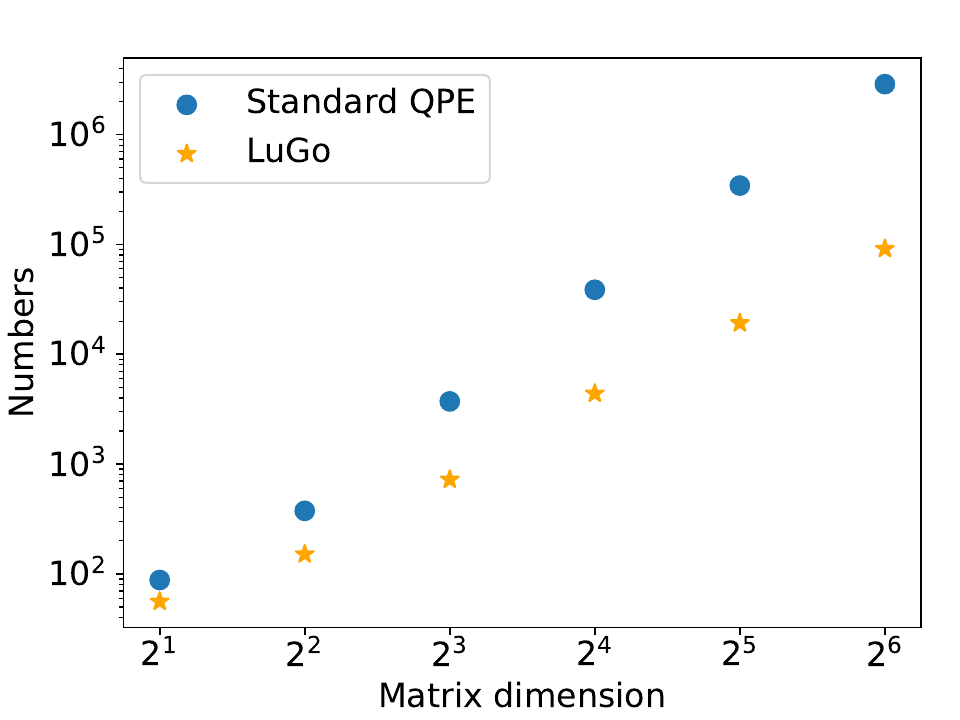}
   \caption{Circuit gate count}
   \label{fig:depth}
 \end{subfigure}
 \begin{subfigure}{.495\textwidth}
   \centering
   \includegraphics[width=0.9\linewidth]{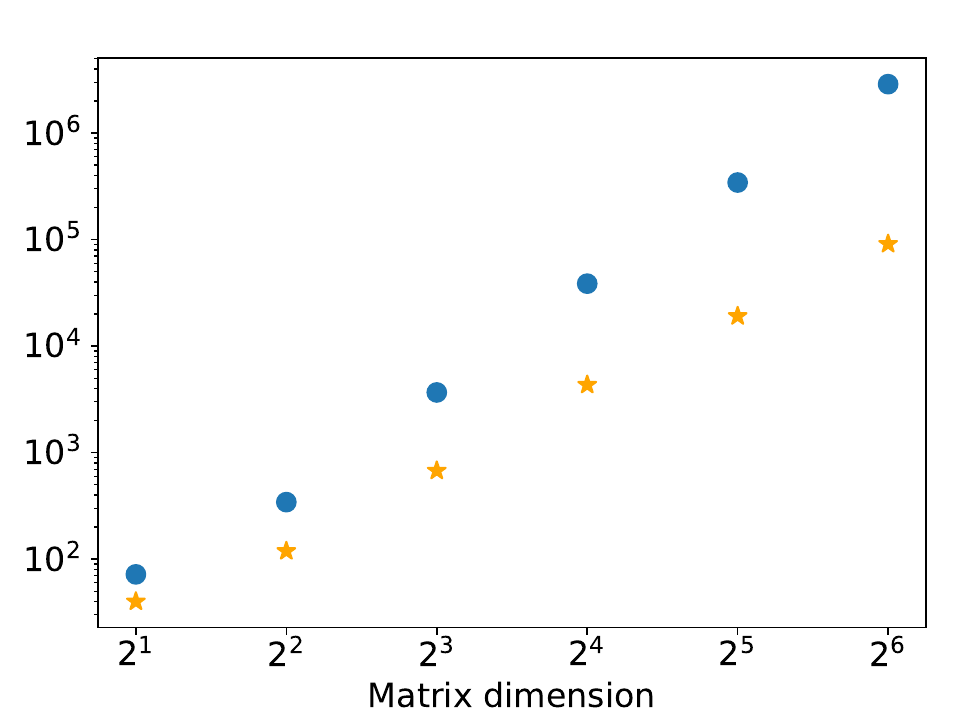}
   \caption{Circuit depth}
   \label{fig:gates}
 \end{subfigure}
 \caption{Circuits statistics comparison of standard QPE and LuGo to generate HHL circuits. From the two figures, LuGo reduced the quantum gates and circuit depth requirement to generate HHL circuits exponentially, providing better scalability over the standard QPE generation method.}
 \label{fig:statistics_HHL}
\end{figure*}

The evaluation is primarily focused on two metrics: execution time and circuit statistics, including gate count and circuit depth.
Fig.~\ref{fig:time_circ_gen} illustrates the time consumption statistics associated with generating HHL circuits using LuGo compared to the standard QPE method. Fig. \ref{fig:time_qpe} shows the time consumption of generating QPE circuits with respect to the size of the unitary matrix input. Fig. \ref{fig:time_qpe+iqpe} presents the time required for generating both QPE and inverted QPE circuits, which account for over $99\%$ of the total circuit generation time. 
Fig.~\ref{fig:time_other_components} details the time consumption involved in generating other components of HHL circuits. These include the HHL circuits' components such as state preparation, eigenvalue reciprocal computation, and controlled rotation circuits. Fig~\ref{fig:time_save} depicts the time required to save the circuits in a {\tt qpy} format file. LuGo demonstrates advantages over standard QPE due to its lower memory storage requirements and more efficient input/output writing processes because of a compact controlled unitary circuit generation approach.

Fig. \ref{fig:time_circ_gen} also shows that the logarithmic scale plots of the LuGo results have a significantly lower slope than the standard QPE method, indicating superior computational efficiency in circuit generation using LuGo. The slopes of the results in Fig. \ref{fig:time_qpe} and \ref{fig:time_qpe+iqpe} are similar, which are $0.75$ for LuGo and $1.01$ for standard QPE. The slope for other components of LuGo is $0.25$ while the slope for standard QPE is $0.74$. As for saving the generated circuit to storage, the slope for LuGo is $0.41$, and for standard QPE is $0.90$.

\begin{figure}
  \centering
  \includegraphics[width=\columnwidth]{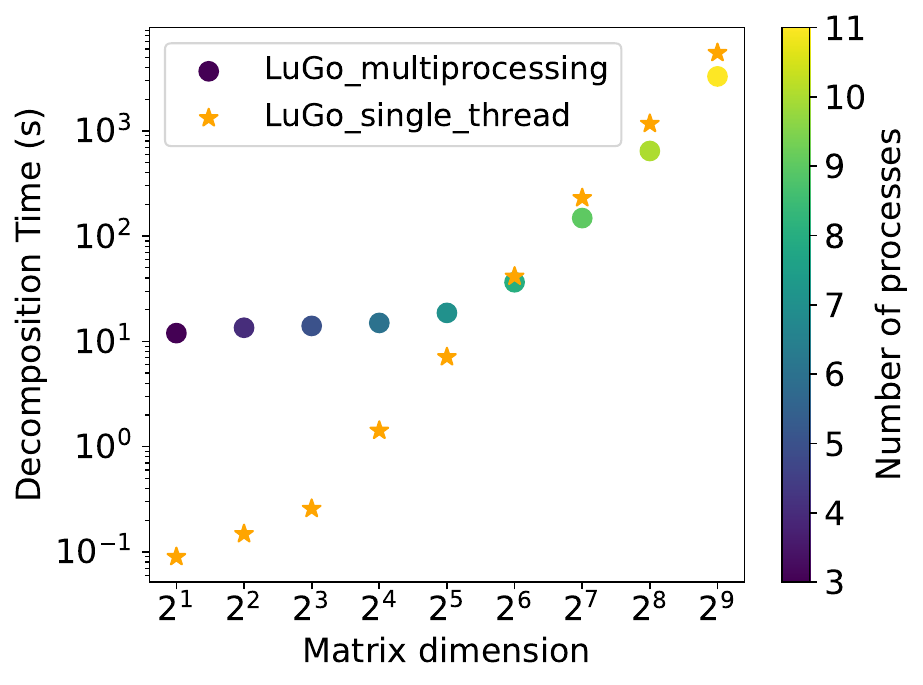}
  \caption{Time consumption for generating HHL circuits using LuGo. Single threaded LuGo has better performance on matrix dimension less than $2^6\times 2^6$ and multithreaded LuGo spent less on matrices larger than $2^7\times 2^7$. The number of processes equals to the number of clock qubits for the problem. For the tridiagonal Topelitz, it requires 3 processes/threads for matrix dimension $2^1$ and 11 processes/threads for matrix dimension $2^9$.}
  \label{fig:luGo_multithread}
\end{figure}

Fig. \ref{fig:transpilation} illustrates the time required for unitary circuit decomposition with respect to the total time consumption in both the standard QPE and the LuGo approaches. The standard QPE requires relatively less time to perform the unitary matrix decomposition compared with LuGo, as the standard QPE requires the decomposition of only the first unitary matrix. However, the subsequent computation requires the duplication of the decomposed first unitary circuit to generate other controlled unitary circuits, leading to worse scaling performance with respect to total time consumption to generate the entire QPE (in our case, HHL) circuit. For LuGo, even though the unitary matrix decomposition time attributes to half of the total time consumption when the matrix dimension is larger than 4; it has better scalability compared to the standard QPE approach for the total time to generate the full circuit. Moreover, LuGo can generate more compact circuits with fewer gates (discussed next in Fig. \ref{fig:statistics_HHL}) and can be computed in an embarrassingly parallel approach with better scalability (results discussed later in Fig. \ref{fig:luGo_multithread}).

Fig~\ref{fig:statistics_HHL} depicts circuit statistics, including total gate count and circuit depth, to generate HHL circuits. LuGo achieves less gates and circuit depth requirement compared with the standard QPE generation approach. LuGo obtained around $1.5\times$ to $2\times$ gate counts and circuit depth reduction for a $2\times 2$ matrix input and over $30\times$ gate and circuit depth reduction over standard QPE for the largest system of equations tested, demonstrating better scalability over standard QPE approach. 

\begin{table*}[t]
  \centering
  \caption{Comparative analysis of time consumption, gate count, and circuit depth between standard QPE and LuGo across various matrix sizes.}\label{tab:HHLcomp}
  \scalebox{1}{
  \begin{tabular}{c|c|ccc|ccc|ccc}
\hline
\multirow{2}{*}{\begin{tabular}[c]{@{}c@{}}Matrix\\ Size\end{tabular}} & \multirow{2}{*}{Qubits} & \multicolumn{3}{c|}{Standard QPE} & \multicolumn{3}{c|}{LuGo}    & \multicolumn{3}{c}{Reduction} \\ \cline{3-11} 
 & & Time (s)   & Gate & Depth    & Time (s) & Gate    & Depth   & Time     & Gate     & Depth   \\ \hline
$2^1\times 2^1$&5 & 0.05 & 88   & 72   & 0.09 & 56   & 40   & 0.51$\times$  & 1.57$\times$  & 1.80$\times$ \\
$2^2\times 2^2$&7 & 0.16 & 375  & 343  & 0.11 & 151  & 119  & 1.48$\times$  & 2.48$\times$  & 2.88$\times$ \\
$2^3\times 2^3$& 9 & 1.63 & 3,718 & 3,672 & 0.34 & 722  & 676  & 4.86$\times$  & 5.15$\times$  & 5.43$\times$ \\
$2^4\times 2^4$& 11 & 17.37   & 38,633   & 38,571   & 1.34 & 4,388 & 4,326 & 12.95$\times$ & 8.80$\times$  & 8.92$\times$ \\
$2^5\times 2^5$ & 13 & 189.26  & 343,250  & 343,170  & 6.50 & 19,194   & 19,114   & 29.13$\times$ & 17.88$\times$ & 17.95$\times$   \\
$2^6\times 2^6$& 15 & 1908.18 & 2,876,007 & 2,875,907 & 37.65   & 91,054   & 90,954   & 50.68$\times$ & 31.59$\times$ & 31.62$\times$   \\
$2^7\times 2^7$&17 & Timeout  & Timeout  & Timeout  & 148.27  & 416,462  & 416,340  & N/A   & N/A   & N/A  \\
$2^8\times 2^8$&19 & Timeout  & Timeout  & Timeout  & 643.42  & 1,866,991 & 1,866,845 & N/A   & N/A   & N/A  \\
$2^9\times 2^9$&21 & Timeout  & Timeout  & Timeout  & 3287.47 & 8,257,810  & 8,257,638  & N/A   & N/A   & N/A  \\ \hline
\end{tabular} 
}
  \end{table*}

\begin{table*}[t]
  \centering
  \caption{Comparative analysis of U3 and CX gates between standard QPE and LuGo across various matrix sizes.}\label{tab:decomposition}
  \scalebox{1}{
\begin{tabular}{c|c|cc|cc|cc}
\hline
\begin{tabular}[c]{@{}c@{}}Matrix\\ Size\end{tabular} & Qubits & \multicolumn{2}{c|}{Standard QPE} & \multicolumn{2}{c|}{LuGo} & \multicolumn{2}{c}{Reduction} \\ \hline
Gates   &  & U3   & CX   & U3  & CX  & U3  & CX  \\ \hline
$2^1\times 2^1$   & 5  & 131  & 108  & 113   & 88  & 1.16$\times$  & 1.23$\times$  \\
$2^2\times 2^2$   & 7  & 1,215   & 880  & 418   & 310   & 2.91$\times$  & 2.84$\times$  \\
$2^3\times 2^3$   & 9  & 16,121  & 11,968  & 2,506  & 1,741  & 6.43$\times$  & 7.04$\times$  \\
$2^4\times 2^4$   & 11   & 170,807   & 127,612   & 16,156   & 11,067   & 10.57$\times$ & 11.53$\times$ \\
$2^5\times 2^5$   & 13   & 1,515,700  & 1,137,060  & 71,681   & 48,868   & 21.15$\times$ & 23.27$\times$ \\
$2^6\times 2^6$   & 15   & 12,712,831   & 9,561,728  & 342,903  & 233,391  & 37.07$\times$ & 40.97$\times$ \\
$2^7\times 2^7$   & 17   & Timeout  & Timeout & 1,572,281   & 1,070,837   & N/A   & N/A   \\
$2^8\times 2^8$   & 19   & Timeout  & Timeout & 7,055,284   & 4,807,733   & N/A   & N/A   \\
$2^9\times 2^9$   & 21   & Timeout  & Timeout & 31,190,810  & 21,262,808  & N/A   & N/A   \\ \hline
\end{tabular}
}
  \end{table*}

  \begin{figure*}[bt!]
 \begin{subfigure}{.48\textwidth}
   \centering
   \includegraphics[width=0.9\linewidth]{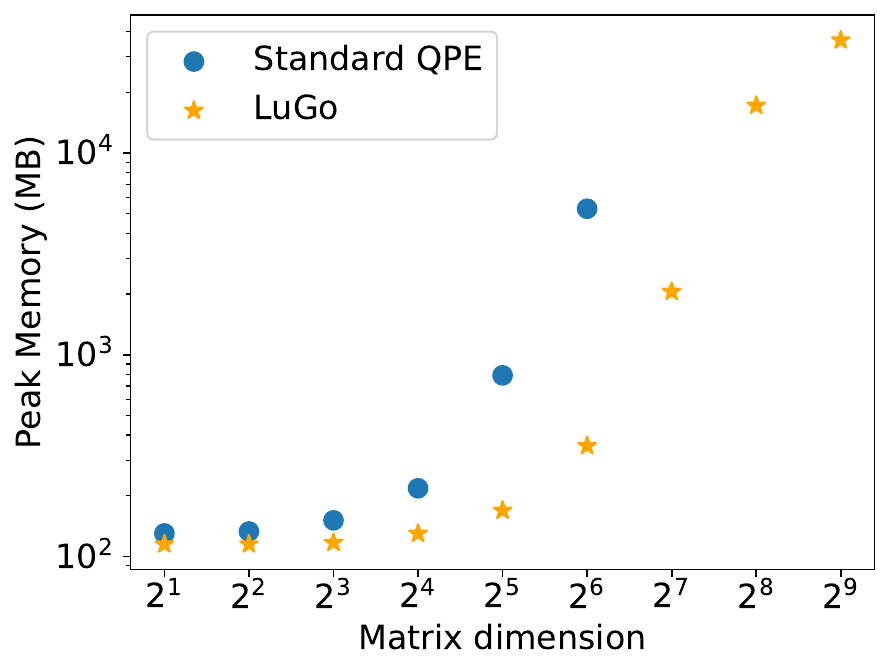}
   \caption{Peak memory consumption using standard QPE and LuGo.}
   \label{fig:peak_mem}
 \end{subfigure}
 \begin{subfigure}{.48\textwidth}
   \centering
   \includegraphics[width=0.9\linewidth]{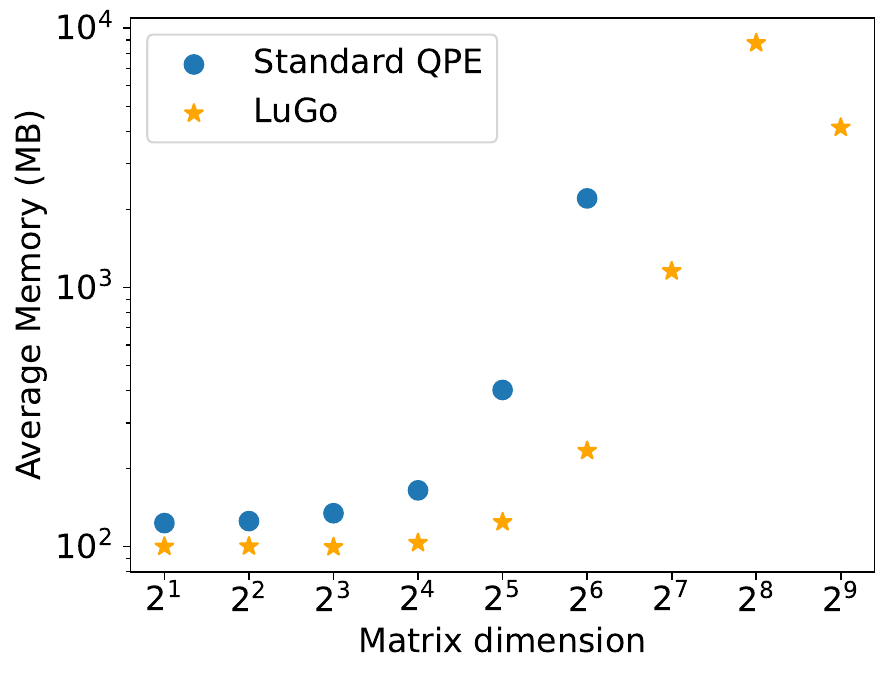}
   \caption{Average memory consumption using standard QPE and LuGo.}
   \label{fig:ave_mem}
 \end{subfigure}
 \caption{Memory footprint of generating HHL circuits using LuGo and standard QPE.}
 \label{fig:memory_footprint}
\end{figure*}
Tab.~\ref{tab:HHLcomp} provides the numeric data of comparing the total time consumption and circuit statistics of generating HHL circuits using the standard QPE circuit generation design and the LuGo framework. The first column indicates the size of the matrix input. It is important to note that the matrix input will always be padded to achieve dimensions of $2^n \times 2^n$ for computational purposes. The second column indicates the total number of qubits required for the HHL circuit. This adjustment ensures compatibility with the requirements of the HHL algorithm, which necessitates matrices of this form. The second through fourth columns display the time consumption and circuit statistics, specifically total gates and circuit depth, using a standard QPE generation method. To collect the mentioned data, we specifically use the {\tt depth} and {\tt count\_ops} functions integrated in Qiskit for the statistics of the generated quantum circuits.
Columns five to seven detail the time consumption and corresponding circuit statistics using LuGo. The time consumption is the least time amongst the single-thread and multithreaded LuGo implementation, later discussed in Fig.~\ref{fig:luGo_multithread}. Both LuGo and the standard QPE method compose a quantum circuit that only consists of pre-defined quantum gates, which contains a basis library of `cu', 'ccx', 'cp', `ry', `cx', `h', `u2', and `p' gates during the circuit generation phase. All unitary gates are decomposed during the circuit generation phase.  Finally, the last three columns illustrate the enhancement achieved by LuGo over a standard QPE implementation in terms of time, gate count, and circuit depth.

The HHL circuit generation algorithm using standard QPE can generate HHL circuits with matrix sizes from $2^1 \times 2^1$ to $2^6\times 2^6$ within two hours using one node of the Frontier supercomputer. Note that the standard QPE implementation is not a parallel implementation. Whereas the LuGo framework can generate circuits for matrix sizes up to $2^9 \times 2^9$, which is a $64\times$ matrix size scalability compared to the existing approach for the given time limit. As for gate counts and circuit depth, LuGo achieves gate count reduction over the standard QPE method from $1.57\times$ with matrix size $2\times 2$ to over $30\times$ with matrix size $2^6\times 2^6$, demonstrating higher scalability on time consumption and circuit statistics. 

To further validate the performance of LuGo, we transpiled the quantum circuit with the integrated function in Qiskit with the default optimization level 2, and we compared the number of U3 and CX gates used in the QPE circuits generated by both LuGo and standard QPE. The results are summarized in Tab.~\ref{tab:decomposition}. The table shows that LuGo consistently reduces the number of U3 and CX gates across all matrix sizes compared to the standard QPE method. The reduction factor ranges from $1.16\times$ to $37.07\times$ for U3 gates and from $1.23\times$ to $40.97\times$ for CX gates, demonstrating significant improvements in gate efficiency.

 \begin{figure*}[t!]
 \begin{subfigure}{.48\textwidth}
   \centering
   \includegraphics[width=\linewidth]{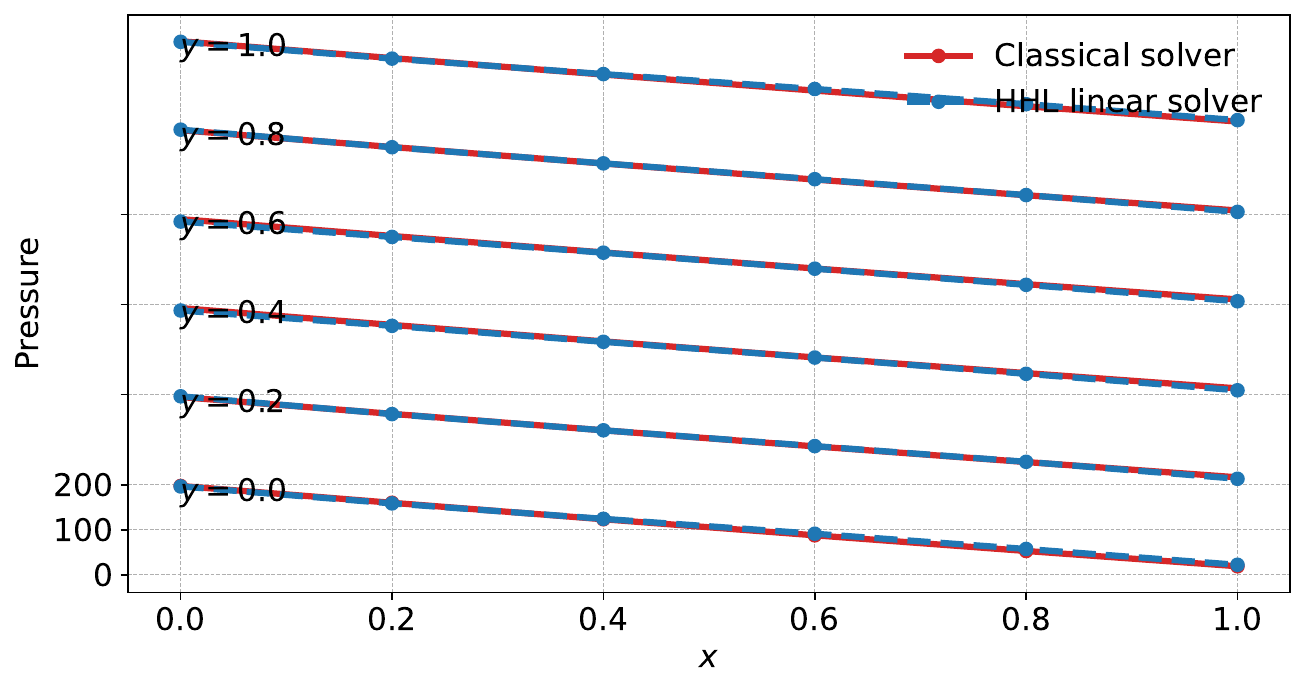}
\end{subfigure}
 \begin{subfigure}{.48\textwidth}
   \centering
   \includegraphics[width=\linewidth]{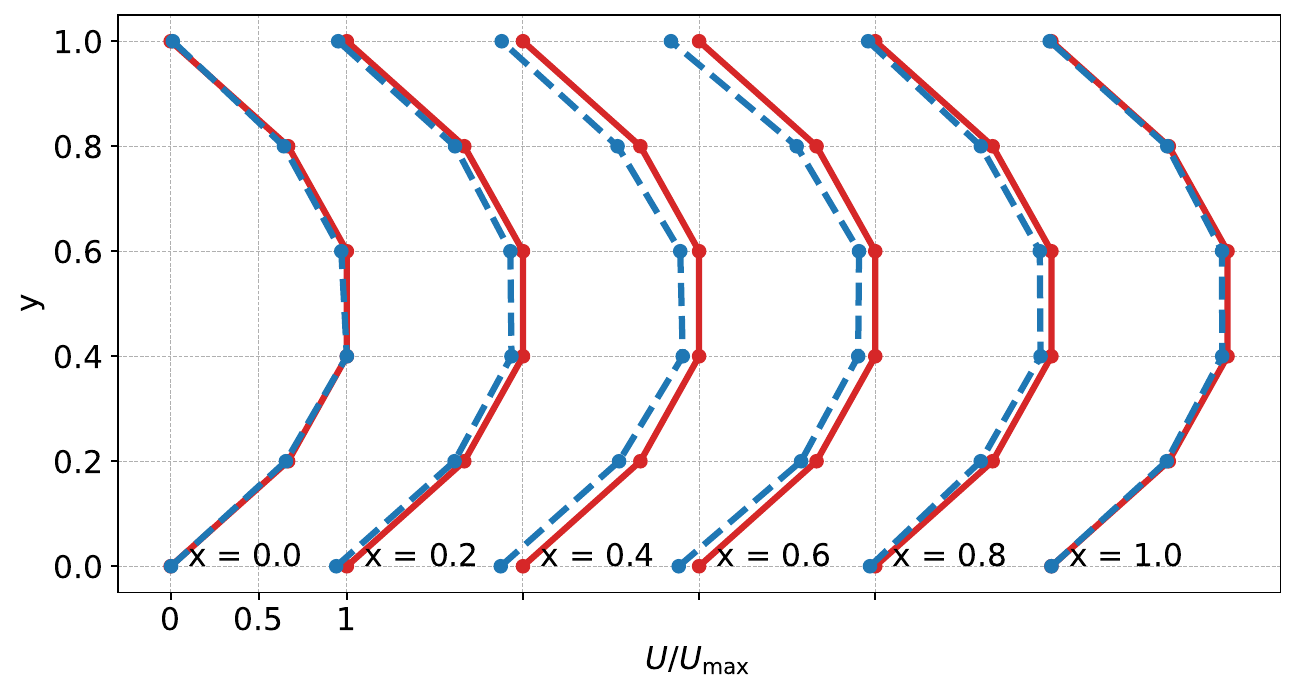}
\end{subfigure}
 \caption{(a) Pressure and (b) velocity profiles of the 2D Hele--Shaw flow solved using HHL algorithm empowered by LuGo QPE.}
 \label{fig:2d_Hele_Shaw}
\end{figure*}

As discussed in Section \ref{sec:HHL}, the generation of HHL circuit requires both QPE and inverse QPE circuits. Traditionally, the generation of the inverse circuit is to simply reverse the order of the circuits and take the inverse operation of each gate of the original circuit. However, such methods are not optimized for multi-threading processes. For computations that requires both QPE and inversed QPE circuit computation, we parallelized the computation of inversed QPE by generating both unitary and inversed unitary circuits to improve the overall performance. 

Fig. \ref{fig:luGo_multithread} illustrates the comparison of time consumption between the single-threaded and multithreaded versions of the LuGo framework. The number of threads used by LuGo equals to the number of clock qubits. The number of clock qubits that HHL algorithm requires are related to the dimension and condition number of the input matrix \cite{HHL}. For tridiagonal Toeplitz matrix, the number of clock bits equals to $\log_2 N + 2$, where $N$ denotes for the dimension of the input matrix. The initialization of the Python multithreading package {\tt ray} introduces additional overhead, causing all test cases to exceed 10 seconds for completion. Thus, for matrices with dimensions up to $2^6 \times 2^6$, the single-threaded version demonstrates superior performance since it avoids the overhead associated with initializing the multithreading package. However, for matrix sizes exceeding $2^6 \times 2^6$, the multithreaded version of LuGo outperforms its single-threaded counterpart. For matrix size of $2^9\times 2^9$, LuGo multithreaded version takes $3287.47$ seconds, and the single-thread version requires $5504.17$ seconds to perform the task, which is a $1.67\times$ reduction on time consumption using the multithreaded version.

To emphasize the memory efficiency of LuGo, we monitored the memory usage during the execution of both LuGo and standard QPE methods using the {\tt psutil} Python package. Fig.~\ref{fig:memory_footprint} illustrates the peak and average memory consumption for generating HHL circuits using both methods. The results indicate that LuGo consistently consumes less memory than the standard QPE approach across all tested matrix sizes. Specifically, LuGo uses approximately $1.13\times$ to $14.98\times$ less peak memory and $1.23\times$ to $9.42\times$ less average memory compared to the standard QPE method. This reduction in memory usage is particularly significant for larger matrix sizes, where efficient memory management becomes crucial for successful circuit generation.

\begin{figure}
 \centering
 \includegraphics[width=\columnwidth]{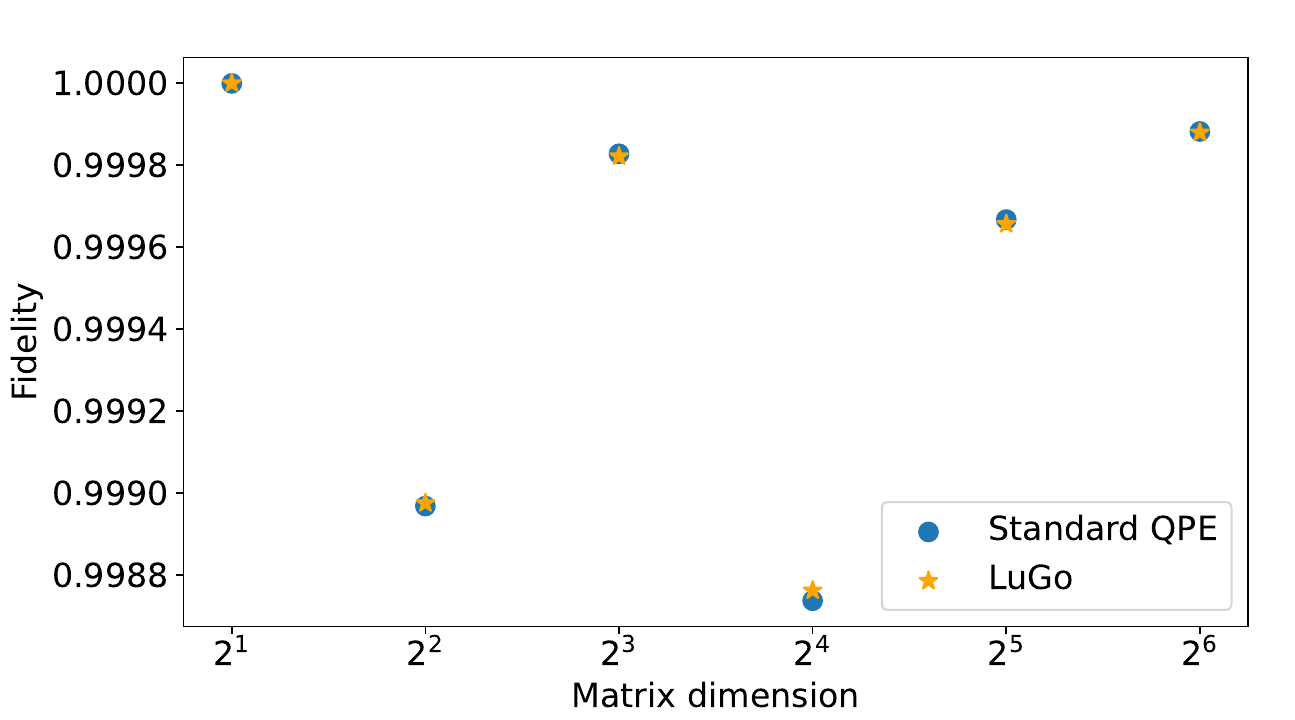}
 \caption{Fidelity comparison of tridiagonal Toeplitz linear systems of equations. Fidelity is defined by the difference between the classical linear solver solution ({\tt numpy} linear solver Python package) and the quantum solution using the HHL algorithm. A fidelity of $1$ means no difference, while a low score close to $0$ suggests discrepancies between the estimated and true values.}
 \label{fig:fidelity}
\end{figure}

We also compare the fidelity results of LuGo against the standard QPE across varying tridiagonal Toeplitz matrix sizes ranging from $2^1 \times 2^1$ to $2^6 \times 2^6$. The circuits are run using an ideal quantum simulator, Qiskit's {\tt AerSimulator}, with $1,000,000$ shots for each circuit for $50$ repetitions. The results are averaged for each test case to eliminate the randomness of each simulation run. The results in Fig.~\ref{fig:fidelity} demonstrate remarkable consistency in performance between both approaches, with all computed fidelity exceeding $0.998$ across the tested matrix sizes. Notably, there is no discernible difference in the fidelity outcomes between LuGo and standard QPE, highlighting the robustness and reliability of LuGo. This similarity in performance underscores the validity of LuGo as an effective alternative to the standard QPE for phase estimation tasks, even as the complexity of the problem increases.

To demonstrate the versatility of LuGo, we adopt a canonical two-dimensional (2D) Hele--Shaw fluid flow problem \cite{gopalakrishnan2024solving}, with the results shown in Fig.~\ref{fig:2d_Hele_Shaw}. Within 600 seconds run time on Frontier within one node, the pressure and velocity profile are generated and the entire process is complete, including the initialization of matrix (non-Hermitian and size $36\times 36$, which is not a power of 2) and vector for the Hele--Shaw problem, pre-processing of matrix and vector input for HHL algorithm compatibility, HHL circuit generation enabled by LuGo QPE, circuit transpilation and simulation using Qiskit's {\tt AerSimutor}, and post-processing the results. The generated HHL circuit for the pressure solver contains 20 qubits, a total of 565,789 gates with a circuit depth of 565,585 for {\tt AerSimulator}. The HHL circuit solving the velocity profile requires 20 qubits and 565,781 gates with a circuit depth of 565,577. The quantum result using $10^6$ shots has a fidelity of $99.96\%$ and $99.24\%$ for pressure and velocity, respectively, compared with the analytical solver. The generated Hele--Shaw problem is impossible to solve on the Frontier supercomputer within 2 hours wall time if using the standard QPE approach of the HHL algorithm. The results demonstrate the capability of LuGo QPE to solve arbitrary linear system of equations with faster quantum circuit generation and simulation compared to the standard QPE approach. \section{Discussion and Conclusion}\label{sec:Conclusion}
We introduce the LuGo framework that enhances the performance of Quantum Phase Estimation (QPE) by reducing the time to generate the circuit with a more compact QPE circuit design. LuGo can apply on any quantum circuits that utilize QPE circuit. We demonstrate the capability of LuGo by generating circuits for a quantum linear systems algorithm, the HHL algorithm. The LuGo algorithm achieves $50.68\times$ reduction on time consumption and over $31\times$ on quantum gates and circuit depth reduction to solve a system of linear equations with a matrix size of $2^6 \times 2^6$. LuGo can generate HHL circuits for solving a matrix size of $2^9 \times 2^9$ within one hour on the Frontier supercomputer, which is a $64\times$ scalability on matrix size over the standard QPE approach. By simulating the generated linear solver circuits, LuGo obtained almost identical fidelity compared to standard QPE circuit generation method. We also performed a two-dimentional Hele--Shaw problem with resolution where using the standard approach is infeasible. 

Our LuGo framework demonstrates better scalability on circuit generation time consumption, gate counts, and circuit depth over the existing approach. With the tremendous efficiency improvement of QPE circuit generation, LuGo achieves one more step towards real-world applications of quantum computing. However, the scalability of the QPE algorithm is still limited due to heavy computation overhead and generation of controlled-unitary circuits with increase in the number of clock qubits. The controlled-unitary circuits highly rely on unitary circuits decomposition algorithms, which still lack scalability and limits the performance of QPE circuits generation algorithm. Therefore, it is still crucial to improve the controlled-unitary circuits decomposition algorithms to use fewer gates and reduced time consumption while maintaining high fidelity for the efficient QPE computation.

\section*{Acknowledgement}
This research used resources of the Oak Ridge Leadership Computing Facility at the Oak Ridge National Laboratory, which is supported by the Office of Science of the U.S. Department of Energy under Contract No. DE-AC05-00OR22725. The authors sincerely appreciate the insightful discussions and feedback from Alessandro Baroni, Seongmin Kim, Antigoni Georgiadou, In-Saeng Suh, Tom Beck, and Travis Humble.

\section*{Data and code availability}
Correspondence and requests for material should be addressed to Chao Lu and Muralikrishnan Gopalakrishanan Meena. Our codes are available for academic or commercial use. Please reach out to \href{luc1@ornl.gov}{luc1@ornl.gov}, \href{gopalakrishm@ornl.gov}{gopalakrishm@ornl.gov}, or \href{partnerships@ornl.gov}{partnerships@ornl.gov}. 

\balance
\bibliographystyle{elsarticle-num} 
\bibliography{references}
\end{document}